\documentclass[11pt, a4paper]{article}\usepackage[]{graphicx}\usepackage[]{color}
\makeatletter
\def\maxwidth{ %
  \ifdim\Gin@nat@width>\linewidth
    \linewidth
  \else
    \Gin@nat@width
  \fi
}
\makeatother

\definecolor{fgcolor}{rgb}{0.345, 0.345, 0.345}
\newcommand{\hlnum}[1]{\textcolor[rgb]{0.686,0.059,0.569}{#1}}%
\newcommand{\hlstr}[1]{\textcolor[rgb]{0.192,0.494,0.8}{#1}}%
\newcommand{\hlcom}[1]{\textcolor[rgb]{0.678,0.584,0.686}{\textit{#1}}}%
\newcommand{\hlopt}[1]{\textcolor[rgb]{0,0,0}{#1}}%
\newcommand{\hlstd}[1]{\textcolor[rgb]{0.345,0.345,0.345}{#1}}%
\newcommand{\hlkwb}[1]{\textcolor[rgb]{0.69,0.353,0.396}{#1}}%
\newcommand{\hlkwc}[1]{\textcolor[rgb]{0.333,0.667,0.333}{#1}}%
\newcommand{\hlkwd}[1]{\textcolor[rgb]{0.737,0.353,0.396}{\textbf{#1}}}%

\usepackage{framed}
\makeatletter
\newenvironment{kframe}{%
 \def\at@end@of@kframe{}%
 \ifinner\ifhmode%
  \def\at@end@of@kframe{\end{minipage}}%
  \begin{minipage}{\columnwidth}%
 \fi\fi%
 \def\FrameCommand##1{\hskip\@totalleftmargin \hskip-\fboxsep
 \colorbox{shadecolor}{##1}\hskip-\fboxsep
     \hskip-\linewidth \hskip-\@totalleftmargin \hskip\columnwidth}%
 \MakeFramed {\advance\hsize-\width
   \@totalleftmargin\z@ \linewidth\hsize
   \@setminipage}}%
 {\par\unskip\endMakeFramed%
 \at@end@of@kframe}
\makeatother

\definecolor{shadecolor}{rgb}{.97, .97, .97}
\definecolor{messagecolor}{rgb}{0, 0, 0}
\definecolor{warningcolor}{rgb}{1, 0, 1}
\definecolor{errorcolor}{rgb}{1, 0, 0}
\newenvironment{knitrout}{}{} % an empty environment to be redefined in TeX

\usepackage{alltt}

\usepackage{verbatim}
\usepackage[vmargin=3cm, hmargin=2cm]{geometry}
\usepackage{url}
\usepackage{hyperref}
\usepackage{fancyhdr}
\usepackage{color}
\usepackage{amsmath, amssymb}
\usepackage{longtable}
\usepackage{lscape}
\usepackage[super, comma,numbers]{natbib}
\usepackage{xspace}
\usepackage{booktabs}
\usepackage{newfloat}
\usepackage{rotating}
\usepackage{scrextend}
\usepackage{csquotes}
\DeclareFloatingEnvironment[
    placement=!htbp
]{listing}
\usepackage{fancyvrb}
\usepackage{lmodern}
\usepackage{nameref}
\usepackage[T1]{fontenc}
\usepackage{listings}
\usepackage[font=sf, labelfont={sf}, margin=1cm]{caption}

\usepackage[latin1]{inputenc}

\IfFileExists{upquote.sty}{\usepackage{upquote}}{}
\begin{document}

\begin{center}
    \Large
    \textbf{Meta-analysis of few studies involving rare events}

    \vspace{0.4cm}
    \large
    \textbf{Burak K\"ursad G\"unhan},\footnote{\textit{Department of Medical Statistics, University Medical Center G\"ottingen, G\"ottingen, Germany} \label{goettingen}} \footnote{\textit{Correspondence to: Burak K\"ursad G\"unhan; email: \texttt{burak.gunhan@med.uni-goettingen.de}}} \textbf{Christian R\"over},\footref{goettingen} \textbf{Tim Friede}\footref{goettingen} 
    \vspace{0.9cm}
\end{center}

Meta-analyses of clinical trials targeting rare events face particular challenges when the data lack adequate numbers of events for all treatment arms. Especially when the number of studies is low, standard meta-analysis methods can lead to serious distortions because of such data sparsity. To overcome this, we suggest the use of \emph{weakly informative priors} (WIP) for the treatment effect parameter of a Bayesian meta-analysis model, which may also be seen as a form of penalization. As a data model, we use a binomial-normal hierarchical model (BNHM) which does not require continuity corrections in case of zero counts in one or both arms. We suggest a normal prior for the log odds ratio with mean 0 and standard deviation 2.82, which is motivated (1) as a symmetric prior centred around unity and constraining the odds ratio to within a range from 1/250 to 250 with 95\% probability, and (2) as consistent with empirically observed effect estimates from a set of \mbox{$37\,773$} meta-analyses from the Cochrane Database of Systematic Reviews. In a simulation study with rare events and few studies, our BNHM with a WIP outperformed a Bayesian method without a WIP and a maximum likelihood estimator in terms of smaller bias and shorter interval estimates with similar coverage. Furthermore, the methods are illustrated by a systematic review in immunosuppression of rare safety events following paediatric transplantation. A publicly available \textbf{R} package, \texttt{MetaStan}, is developed to automate the \textbf{Stan} implementation of meta-analysis models using WIPs.\\

\textbf{Keywords:} Random effects meta-analysis, rare events, few studies, Bayes, weakly informative priors

% =======Some Initial settings for R==============

%========End of Initial Settings for R============

%=================================================
\section{Introduction}\label{sec:intro}
%=================================================
Individual clinical studies are often underpowered to detect differences in the rates of rare events, for example safety events, and thus meta-analysis may be the only way to obtain reliable evidence of treatment differences with regard to the rare events.\cite{higginscochrane} On the other hand, meta-analysis of clinical studies for rare events faces particular challenges, since the numbers of events might be very small in some treatment arms. The problem is even more pronounced when some studies have no events either in one or both treatment arms (so-called \emph{single-zero} or \emph{double-zero} studies).

Standard (approximate) meta-analysis methods, for example the normal-normal hierarchical model,\cite{Hedges1985189} require a \emph{continuity correction} in case of single-zero or double-zero studies, that is, the addition of a fixed value (typically 0.5) to all cells of the contingency table for studies with no events. Such simple approaches have been found problematic for meta-analyses involving rare events.\cite{SIM:SIM2528} Therefore, statistical models based on exact distributional assumptions have been suggested. These include different parametrizations of the binomial-normal hierarchical model,\cite{SIM:SIM7588} a mixed effects conditional logistic model,\cite{SIM:SIM4040} a Poisson-normal hierarchical model,\cite{BIMJ:BIMJ1584} a Poisson-Gamma hierarchical model,\cite{SIM:SIM3964} or a beta-binomial model.\cite{SIM:SIM6383} In this paper, we focus on a parametrization of the binomial-normal hierarchical model (BNHM) which was suggested by Smith et al.\cite{SIM:SIM4780142408} 

Consider an extreme case of meta-analysis of rare events, where all studies include no events for the same treatment arm. This data sparsity problem in a meta-analysis can be seen as a \emph{separation} problem in the logistic regression context \cite{albert1984} in which case maximum likelihood estimate (MLE) for treatment effect parameter does not exist. A very useful way to deal with separation problems, or, more generally, data sparsity in logistic regression is \emph{penalization}, that is, adding a penalty (adjustment) term to the original likelihood function to regularize (or stabilize) the estimates.\cite{greenland2015penalization} In a frequentist framework, penalty terms may be specified so that these ``nudge'' the MLE into a desired direction if the maximum is not or poorly defined; one such example is Firth penalization.\cite{Firth1993,Heinze2002, greenland2015penalization} From a Bayesian viewpoint, penalization may often be motivated as \emph{weakly informative priors} (WIPs) that are multiplied to the likelihood function.\cite{gelman2008} 

Numbers of studies included in meta-analyses are typically small, posing additional challenges.\cite{Davey2011} For Bayesian meta-analysis of few studies, different WIPs have been suggested for the heterogeneity parameter, see \cite{SIM:SIM5821} for penalized MLE approach, and also see \cite{gelman2006,JRSM:JRSM1217,BIMJ:BIMJ1725,williams2018bayesian, Benderetal2018} for fully-Bayesian inference. Here we consider the meta-analysis of few studies targeting rare events. To deal with data sparsity present in the meta-analysis of few studies with rare events, we suggest the use of WIPs for the treatment effect parameter in a fully-Bayesian context inspired by penalization ideas.\cite{greenland2015penalization,gelman2008} We use a BNHM which is parameterized in terms of baseline risks and a treatment effect for the data. Note that this is a contrast-based model meaning that relative treatment effects are assumed to be exchangeable across trials.\cite{dias2015absolute}
Our suggested default WIP for the treatment effect parameter is motivated via the consideration of the prior expected range of treatment effect values. Furthermore, it is consistent with effect estimates empirically observed in a large set of meta-analyses from the Cochrane Database of Systematic Reviews (CDSR) with binary endpoints.

The main contribution of this paper is the introduction of default WIPs as penalization for treatment effect parameters to deal with data sparsity in the meta-analysis of few studies involving rare events. Another contribution is the introduction of an \textbf{R} package, \textbf{MetaStan} (\href{https://github.com/gunhanb/MetaStan}{https://github.com/gunhanb/MetaStan}), which is developed to automate a Bayesian implementation of meta-analysis models using WIPs as described in the paper, and which is publicly available from Github. In Section~\ref{sec:app}, we describe a systematic review concerning rare safety events associated with immunosuppressive therapy following paediatric transplantation. In Section~\ref{sec:WIP}, we describe the application of weakly informative priors (WIPs) for the treatment effect parameter. We review a BNHM for meta-analysis, discuss the derivation of WIP, and an empirical investigation of treatment effect parameter estimates from the CDSR. Long-run properties of different methods including the proposed one are investigated in the simulation studies in Section~\ref{sec:sim}. In Section~\ref{sec:revisited}, the example is revisited to illustrate the proposed method and its implementation. We close with some conclusions and provide a discussion.

%=================================================
\section{An application in paediatric transplantation} \label{sec:app}
%=================================================#
Several rare paediatric liver diseases can nowadays be successfully treated by liver transplantation with good long-term outcomes. Crins et al\cite{PETR:PETR12362} conducted a systematic review of controlled but not necessarily randomized studies of the Interleukin-2 receptor antibodies (IL-2RA) basiliximab and daclizumab in paediatric liver transplantation. Primary outcomes were acute rejections (ARs), steroid-resistant rejections (SRRs), graft-loss and death. Their analyses were based on a random effects meta-analysis using a restricted maximum likelihood approach (REML).\cite{metafor} Crins et al\cite{PETR:PETR12362} used risk ratios as effect measures, while we use the odds ratios, here. With rare events, however, these should be very similar. Heterogeneity was assessed using Cochrane's Q test.\cite{Higgins557} Secondary outcomes included renal dysfunction and post-transplant lymphoproliferative disease (PTLD). For illustrative purposes, here we focus on death and PTLDs, and these outcomes are displayed in Table~\ref{tab:Data}.

\begin{table}[htb]
\centering
\caption{Data on patient deaths and post-transplant lymphoproliferative disease (PTLD) from the meta-analysis in paediatric transplantation conducted by Crins et al.\cite{PETR:PETR12362}}
\label{tab:Data}
\begin{tabular}{lcccccccc}
  \toprule
   &\multicolumn{4}{c}{\textbf{Outcome: Death}}     & \multicolumn{4}{c}{\textbf{Outcome: PTLD}}  \\
   \cmidrule(rl){2-5} \cmidrule(rl){6-9}
   &\multicolumn{2}{c}{\textbf{Control}}   & \multicolumn{2}{c}{\textbf{Experimental}} &\multicolumn{2}{c}{\textbf{Control}}   & \multicolumn{2}{c}{\textbf{Experimental}}  \\
  	\cmidrule(rl){2-3} \cmidrule(rl){4-5} \cmidrule(rl){6-7} \cmidrule(rl){8-9}  
  	                                       & Events & Total & Events & Total & Events & Total & Events & Total         \\ \midrule
     Heffron et al\cite{heffron2003}       &   3    &  20   &   4    &  61   &   -    &  -    &   -    &  - \\ 
     Schuller et al\cite{SCHULLER20051151} &   -    &  -    &   -    &  -    &   0    &  12   &   0    &  18 \\ 
     Ganschow et al\cite{PETR:PETR371}     &   3    &  54   &   1    &  54   &   0    &  54   &   1    &  54 \\ 
     Spada et al\cite{AJT:AJT1406}         &   3    &  36   &   4    &  36   &   1    &  36   &   1    &  36 \\ 
     Gras et al\cite{LT:LT21397}           &   3    &  34   &   2    &  50   &   -    &  -    &   -    &  - \\ 
   \bottomrule
\end{tabular}
\end{table}

The specific problems with meta-analyses concerning rare events outlined in the introduction are prominent here. Firstly, the numbers of events are very small. For the PTLD dataset, there is one single-zero study and one double-zero study. Secondly, there are few studies available, only 4 for deaths and 3 for PTLD. Empirical event rates are lower in 3 of the 4 experimental groups in the data on patient deaths. For PTLD, the data appear inconclusive for Schuller et al\cite{SCHULLER20051151} and Spada et al,\cite{AJT:AJT1406} and only a single event observed in the experimental group suggests an increased risk in the study by Ganschow et al.\cite{PETR:PETR371}
%=================================================
\section{Weakly informative priors for the treatment effect} \label{sec:WIP}
%=================================================
In this section, we present the usage of weakly informative priors (WIPs) for the treatment effect parameter to conduct random effects meta-analysis of rare events with few studies. As a data model, we review a binomial-normal hierarchical model, and then show how to derive a WIP for a treatment effect parameter. Then, empirical evidence obtained from the Cochrane Database of Systematic Reviews supporting the choice of WIPs is illustrated. 
%=================================================
\subsection{Data model} \label{sec:DataModel}
%=================================================
The binomial-normal hierarchical model (BNHM) has been introduced by Smith et al.\cite{SIM:SIM4780142408}
In the BNHM, for each trial $i \in \{1,\ldots,k\}$ and treatment arm $j \in \{0,1\}$, the event counts $r_{ij}$ are modeled using a binomial distribution, that is, $r_{ij} \sim \text{Bin}(\pi_{ij}, n_{ij})$. 
The logit link is used to transform $\pi_{ij}$ onto the log odds ratio scale where effects can be assumed to be additive 
\begin{align}
\text{logit}(\pi_{ij}) &  =  \mu_{i} + \theta_{i} \, x_{ij} \nonumber \\
\theta_{i}             &  \sim \mathcal{N}(\theta, \tau^2) \label{eq:BNHM}
\end{align}
where $x_{ij}$ is a treatment indicator, namely $+0.5$ = experimental (j = 1) and $-0.5$ = control (j = 0). The $\mu_{i}$ are fixed effects denoting the baseline risks of the event in each study $i$, $\theta$ is the mean treatment effect and $\tau$ is the heterogeneity in treatment effects between trials. The BNHM belongs to the family of generalized linear mixed models (GLMMs); this family also includes models for other types of data including continuous or count outcomes. It is important to note here that by treating the baseline risks $\mu_{i}$ as fixed effects, the analysis effectively stratifies the risk by study, as pooling of risks might compromise the studies' randomization. In this sense it constitutes a contrast-based model.\cite{dias2015absolute} Unlike the normal-normal hierarchical model, the BNHM does not rely on a normal approximation, since it builds on the binomial nature of the data directly.

The BNHM can be fitted using frequentist approaches, for example via maximum likelihood estimation (MLE). \cite{SIM:SIM7588} Alternatively, Bayesian methods are commonly used. In a fully Bayesian approach, prior distributions for parameters $\theta$, $\mu_{i}$ and $\tau$ need to be specified. Note that the parameter $\theta$ is on the log-odds ratio scale whereas $\mu_{i}$ are on the log-odds scale. Baseline risks ($\mu_{i}$) may be seen as intercept parameters in a standard logistic regression model. For $\mu_{i}$, we use a vague normal prior with mean 0 and standard deviation 10, following the recommendation by Gelman et al.\cite{gelman2008} The prior choice for $\theta$ is our main focus and will be discussed in Section~\ref{sec:derive}. The prior choice for the heterogeneity parameter $\tau$, which is a standard deviation parameter, has gained much attention in the literature as discussed in the introduction. Friede et al\cite{JRSM:JRSM1217} have shown that for meta-analysis of few studies, the use of WIPs for $\tau$ displays desirable long-run properties in comparison to frequentist alternatives. Following their suggestions, we use a half-normal prior with scale of 0.5 ($\mathcal{HN}(0.5)$) for $\tau$ which has the median of 0.337 with an upper 95\% quantile of 0.98. Values for $\tau$ of 0.25, 0.5, 1, and 2 represent moderate, substantial, large, and very large heterogeneity.\cite{Spiegelhalter2004bayesian} Thus a $\mathcal{HN}(0.5)$ prior captures heterogeneity values typically seen in meta-analyses of heterogeneous studies and will therefore be a sensible choice in many applications. 

%=================================================
\subsection{Derivation of a weakly informative prior for the treatment effect} \label{sec:derive}
%=================================================
A common prior choice for the treatment effect parameter $\theta$ is a non-informative (vague) prior such as normal distribution with a large variance, for example $\mathcal{N}(0, 100^2)$. One way of constructing a WIP works via consideration of the prior expected range of treatment effect values. \cite{Greenland01062006} Before the derivation of the WIP for treatment effect parameter $\theta$, recall that $\theta$ is on the log odds ratio scale. Thus, a value of $\theta = 0$ means an odds ratio of 1, i.e., ``no effect'', and a value of $\theta = 1$ means that odds differ by a factor (ratio) of $\text{exp}(1) = 2.7$.

We assume a symmetric prior centred around zero, implying equal probabilities for positive or negative treatment effects. Symmetry then implies (on the log odds ratio scale) that
\begin{align}
  P(\theta > q) = P(\theta < -q)
\end{align}
where (on the odds ratio scale)
\begin{align}
  \text{exp}(-q) = \frac{1}{\text{exp}(q)}.
\end{align}

The prior's scale parameter $\sigma_{\text{prior}}$ then may be set such that a priori the odds ratio is with 95\% probability confined to a certain range:
\begin{align}
  P(1/\delta < \text{exp}(\theta) < \delta) = 95\%.
\end{align}

In case of a normal prior with standard deviation $\sigma_{\text{prior}}$, we can then simply specify 
\begin{align}
\sigma_{\text{prior}} = \frac{\text{log}(\delta)}{1.96}\label{eq:WIPprior}.
\end{align} 

We conservatively specify $\delta$ as 250, meaning that we consider it unlikely that odds differ between treatment and control groups by a factor of more than 250. By plugging in this number into \eqref{eq:WIPprior}, we obtain $\sigma_{\text{prior}} = 2.82$.

Another way to motivate the prior standard deviation is by using the idea of \emph{unit information priors}.\cite{doi:10.1080/01621459.1995.10476592, bayesmeta} When the treatment effect parameter is on the log odds ratio scale (as in the BNHM), then the standard error is given by $\sqrt{\frac{1}{a} + \frac{1}{b} + \frac{1}{c}+ \frac{1}{d}}$. Assuming equal allocation and a neutral effect, we can simply set the table allocation to $a = b = c = d = \frac{N}{4}$. Therefore, if we (heuristically) reverse the argument, a prior for the log odds ratio with zero mean and 2.82 standard deviation gives\cite{bayesmeta}
\begin{align}
2.82^2 \approx 8 = \sqrt{\frac{1}{\frac{N}{4}} + \frac{1}{\frac{N}{4}} + \frac{1}{\frac{N}{4}}+ \frac{1}{\frac{N}{4}}}.
\end{align}
Hence $N = 2$. In other words, in terms of prior's effective sample size, the choice of $\sigma_{\text{prior}} = 2.82$ is equivalent to adding 2 patients to the dataset. From this, it follows that $\mathcal{N}(0, 2.82)$ is a reasonable choice as a WIP for $\theta$.

Note also the analogy between this WIP and commonly used continuity corrections: Zero entries in a contingency tables are commonly ``fixed'' by adding a correction term of 0.5 to each table cell of the single-zero or double-zero study, which also amounts to a total of 2 patients ``added'' to the data. Note that the continuity correction adds 2 patients to each single-zero or double-zero study, while the use of WIP is equivalent to adding 2 patients to the whole dataset.
%==============================================================================
\subsection{Empirical evidence supporting the weakly informative prior for the treatment effect} \label{sec:cochrane}
%===============================================================================
For an empirical investigation of the WIP for treatment effect parameter, we consider the meta-analysis datasets archived in the CDSR. All systematic reviews in the CDSR are available on the Cochrane Library website,\cite{cochrane} and personal or institutional access is required. For downloading the data from the CDSR and converting to CSV files, we use the program \texttt{Cochrane\_scraper} (version 1.1.0).\cite{davidaspringate201410782} We were able to access all Cochrane systematic reviews available in March 2018 (CD000004 to CD012788). Meta-analyses were excluded if they included only one study, if the analysis was labeled as a subgroup or sensitivity analysis or there was insufficient information for classification, or if all data within the meta-analysis appeared to be erroneous. Finally, we only consider meta-analyses with dichotomous outcomes. In total \mbox{$37\,773$} meta-analysis datasets from \mbox{$4\,712$} reviews are included. Note that we did not distinguish regarding efficacy or safety analyses.

\begin{figure}
  \centering
  \includegraphics[scale=0.65]{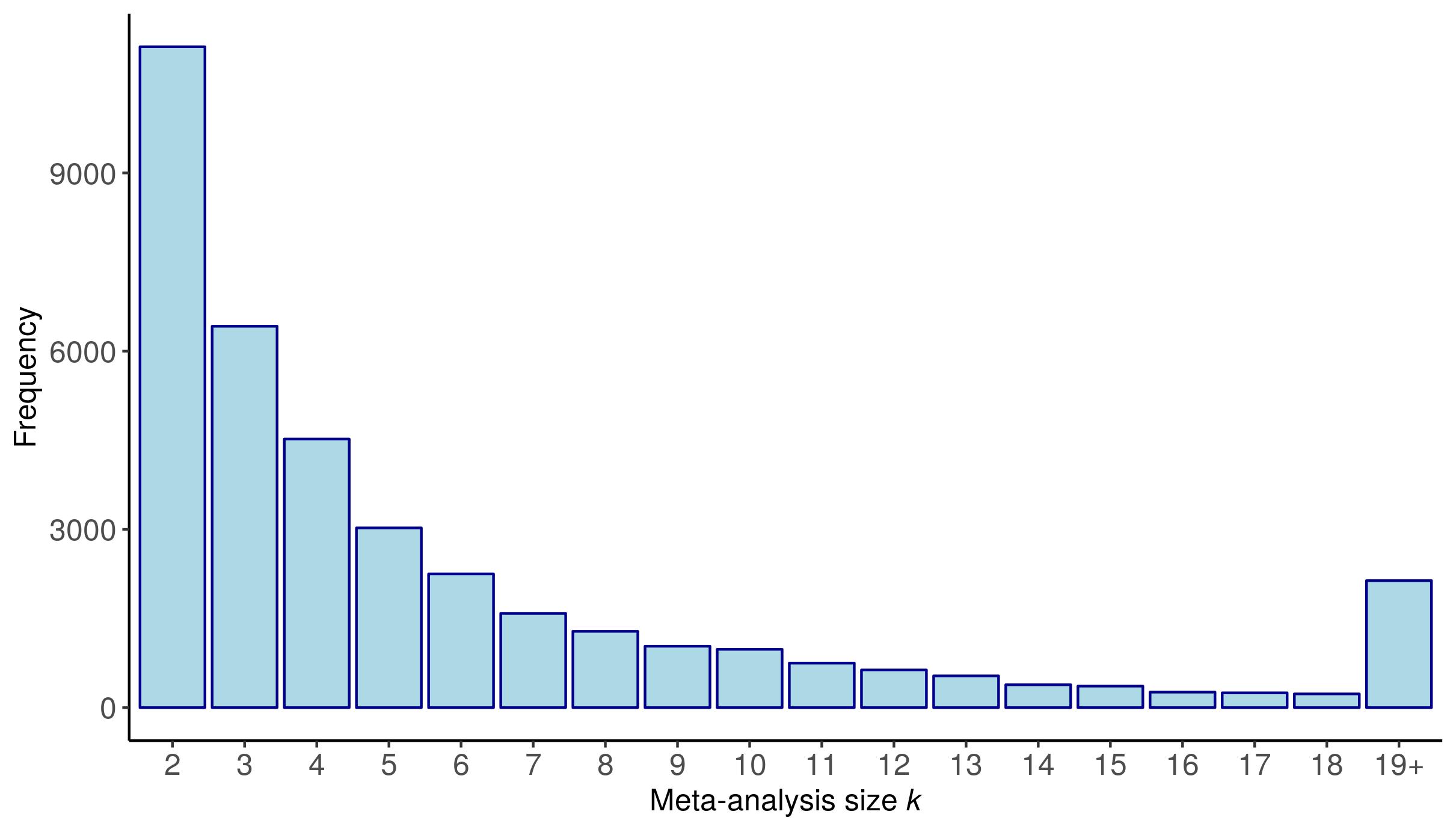}
  \caption{The distribution of numbers of studies included in each meta-analysis obtained from the Cochrane Database of Systematic Reviews (CDSR). The category labelled "19+" corresponds to meta analyses of size 19 or larger.}
  \label{fig:cochrane-masizes}
\end{figure}

The frequency of the number of studies $k$ considered for each meta-analysis is illustrated in Figure 1. The percentage of the meta-analyses including 5 or less studies is 66\%. This figure is consistent with other re-analyses of the CDSR (see e.g.\cite{Davey2011, doi:10.1093/ije/dys041, 10.1371/journal.pone.0069930}). We re-analysed the meta-analysis datasets from the CDSR using the BNHM via a maximum likelihood estimate (MLE) approach. This procedure is implemented using the \textbf{R} package \texttt{lme4}.\cite{lme4} A histogram of the estimates of $\theta$ is illustrated in Figure 2A. 2.5\% and 97.5\% quantiles of the estimates of $\theta$ are -1.94 and 2.06, respectively. By following Turner et al,\cite{doi:10.1093/ije/dys041} we exclude the zero heterogeneity estimates; non-zero estimates of $\tau$ are shown in Figure 2B. The fraction of non-zero heterogeneity estimates is 63\%, which is also consistent with previous findings.\cite{doi:10.1093/ije/dys041} The 95\% quantile of non-zero estimates of $\tau$ is 1.51, while the 95\% quantile of $\tau$ estimates including zeroes is 1.05. The underlying distribution of the estimates of $\theta$ and $\tau$ and their variability are useful to see how large are these estimates in some general population, in this case the CDSR. And, thus these give us a rough sense of what would be a reasonable default prior distribution. Therefore, we suggest the use of WIPs, $\mathcal{N}(0, 2.82)$ for $\theta$ and $\mathcal{HN}(0.5)$ for $\tau$, which are consistent with estimates of $\theta$ and $\tau$ empirically observed in the CDSR.

\begin{figure}
  \centering
  \includegraphics[scale=0.75]{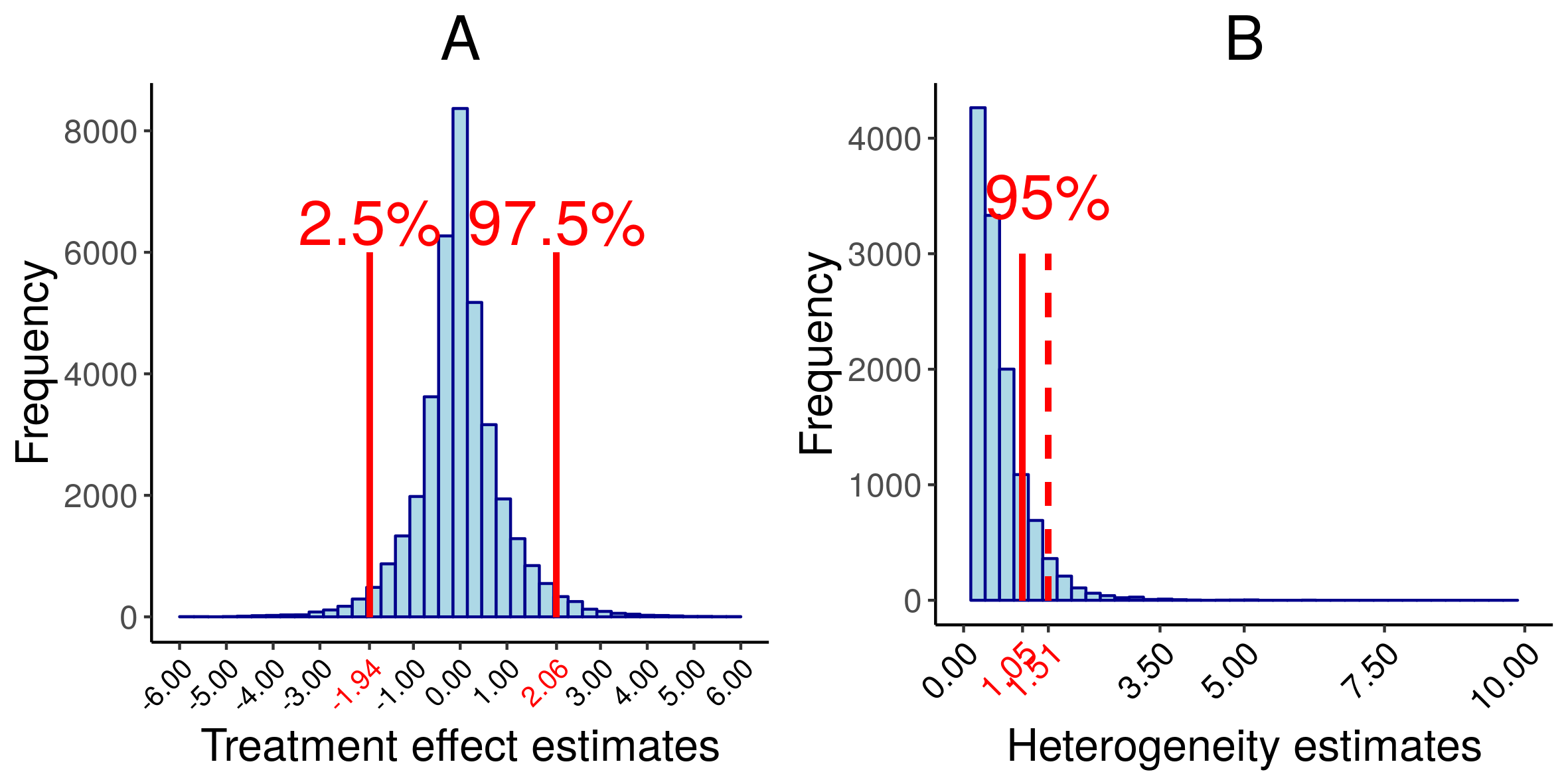}
  \caption{The distribution of the estimates of the mean treatment effect parameter $\hat{\theta}$ (A) and the distribution of the estimates of the (non-zero) heterogeneity standard deviation parameter $\hat{\tau}$ (B) obtained from the re-analysis of meta-analysis datasets in CDSR when the BNHM via MLE is used for estimation. In A, two red lines (-1.94 and 2.06) show the 2.5\% and 97.5\% quantiles of the $\hat{\theta}$, respectively. In B, the solid red line (1.05) and the dashed red line (1.51) indicate the 95\% quantiles of the  $\hat{\tau}$ including zero-estimates and excluding zero-estimates, respectively. The fraction of zero $\hat{\tau}$'s is 63\%.}
  \label{fig:cochrane-d-tau}
\end{figure}

%==============================================================================
\section{Implementation of the proposed procedure in \textbf{R} using \textbf{Stan}} \label{sec:comp}
%===============================================================================
The Bayesian implementation of the BNHM can be fitted with the probabilistic programming language \textbf{Stan}\cite{stan} which employs a modern Markov chain Monte Carlo (MCMC) algorithm, namely Hamiltonian Monte Carlo with the No-U-Turn Sampler.\cite{HMC} It is known that the parametrization of the model can affect the performance of MCMC algorithm. In the presence of sparse data such as in the meta-analysis of few studies involving rare events, Betancourt et al\cite{betancourt2015hamiltonian} showed that non-centred parametrization of a hierarchical model (such as the BNHM) brings computational issues compared to a centred parametrization. Thus, we use the non-centred reparametrized version of the BNHM for our implementations. Specifically, applying both location and scale reparametrization, \eqref{eq:BNHM} becomes $\mu_{i} + \theta_{i} \, x_{ij} + u_{i} \, \tau^2$ where $u_{i} \sim \mathcal{N}(0, 1)$ and $x_{ij} = +0.5$ (experimental) or $x_{ij} = -0.5$ (control).

For practical applications, learning \textbf{Stan}'s syntax  and the required knowledge of available features in the \textbf{Stan} might present a hurdle preventing application of \textbf{Stan}. To this end, we developed a new \textbf{R} package \texttt{MetaStan} which is a purpose-built package defined on top of \texttt{Rstan}, the \textbf{R} interface of the \textbf{Stan}. Our package \texttt{MetaStan} (\href{https://github.com/gunhanb/MetaStan}{https://github.com/gunhanb/MetaStan}) includes the pre-compiled \textbf{Stan} model of the BNHM, which eliminates the compilation time and the need of learning \textbf{Stan}'s syntax. The \textbf{Stan} code for the BNHM is shown in Listing~\ref{fig:Stan_BNHM}. \texttt{MetaStan} includes different options for WIPs of the model parameters of the BNHM. \texttt{MetaStan} syntax is similar to the syntax of the popular meta-analysis package \texttt{metafor}\cite{metafor} so that it should be easy for a \texttt{metafor} user to utilize our package. The syntax of \texttt{MetaStan} is displayed for the paediatric transplantation example in Section~\ref{sec:revisited}, and in Appendix~\ref{app1}, we show how to install and use \texttt{MetaStan}.

\begin{listing}
\begin{Verbatim}[numbers=left,frame=single,fontfamily=courier,fontsize=\footnotesize]
data {
  int<lower=1> N;                           // num studies
  int<lower=0> rctrl[N];                    // num events, control
  int<lower=1> nctrl[N];                    // num patients, control
  int<lower=0> rtrt[N];                     // num events, treatment
  int<lower=1> ntrt[N];                     // num patients, treatment
  vector[2] mu_prior;                       // Prior parameters for mu (mean and stdev)
  vector[2] theta_prior;                    // Prior parameters for theta (mean and stdev)
  real tau_prior;                           // Prior scale for tau
  int tau_prior_dist;                       // Indicator for prior distribution of tau
}

parameters {
  vector[N] mu;                             // baseline risks (log odds)
  real theta;                               // relative trt effect (log OR)
  vector[N] zeta;                           // individual trt effects
  real<lower=0> tau;                        // heterogeneity stdev.
}

transformed parameters {
  real pctrl[N];
  real ptrt[N];

  for(i in 1:N) {
    pctrl[i] = inv_logit(mu[i] - theta * 0.5);
    ptrt[i] = inv_logit(mu[i] + theta * 0.5 + zeta[i] * tau);
  }
}

model {
  // latent variable (random effects)
  zeta ~ normal(0, 1);
  // prior distributions
  mu ~ normal(mu_prior[1], mu_prior[2]);
  theta ~ normal(theta_prior[1], theta_prior[2]);
  if(tau_prior_dist == 1)  tau ~ normal(0, tau_prior)T[0,];
  if(tau_prior_dist == 2)  tau ~ uniform(0, tau_prior);
  if(tau_prior_dist == 3)  tau ~ cauchy(0, tau_prior)T[0,];
  // likelihood
  rctrl~ binomial(nctrl, pctrl);                // control event count
  rtrt ~ binomial(ntrt, ptrt);                  // treatment event count
}
\end{Verbatim}
\caption{\textbf{Stan} code for the binomial-normal hierarchical model.}\label{fig:Stan_BNHM}
\end{listing}

%=================================================
\section{Simulation study} \label{sec:sim}
%=================================================
In order to assess the long-run properties of the proposed approach and compare it to some alternatives, we conducted a simulation study.

%=================================================
\subsection{Simulation setup} \label{sec:setup}
%=================================================
The simulation scenarios are similar to those considered by Friede et al,\cite{JRSM:JRSM1217} but with the important difference that we focus on rare events. The datasets are generated under the BNHM, more specifically \eqref{eq:BNHM}. Numbers of studies ($k \in \{2, 3, 5\}$) and true treatment effects
($\theta = \{-5, -4, -3, -2, -1, -0.5, 0, 0.5, 1, 2, 3, 4, 5\}$) are varied, resulting in a total of 39 simulation scenarios. To reflect the rare-event cases, true baseline risks on the probability scale are taken uniformly  between 0.005 and 0.05. Following Kuss,\cite{SIM:SIM6383} a log-normal distribution is fitted to the sample sizes obtained from the CDSR data, resulting in a log-normal distribution with parameters $\mu=5$ and $\sigma=1$. Hence, sample sizes are generated from $\mathcal{LN}(5, 1^2)$, the minimum sample size is restricted to 2 patients (values below 2 are rounded up to 2), and at least 1 patient in each treatment arm is assumed. The degree of heterogeneity ($\tau$) is taken as $\tau = 0.28$ (moderate heterogeneity), which is the median value of the predictive distribution for between-study heterogeneity in a meta-analysis as estimated by Turner et al.\cite{doi:10.1093/ije/dys041} According to a binomial probability of 0.5, patients were allocated to the treatment groups, thus mimicking randomization. The simulations were carried out with \mbox{10\,000} replications per scenario. The data sparsity is reflected in the average fraction of single-zero or double-zero studies in a simulated meta-analysis dataset which are shown in Figure 3A. Notice that the fractions of the single-zero and double-zero studies are the highest when true treatment effect is -5, and they are decreasing with the increase of the treatment effect.

\begin{figure}
  \centering
   \includegraphics[scale=0.5]{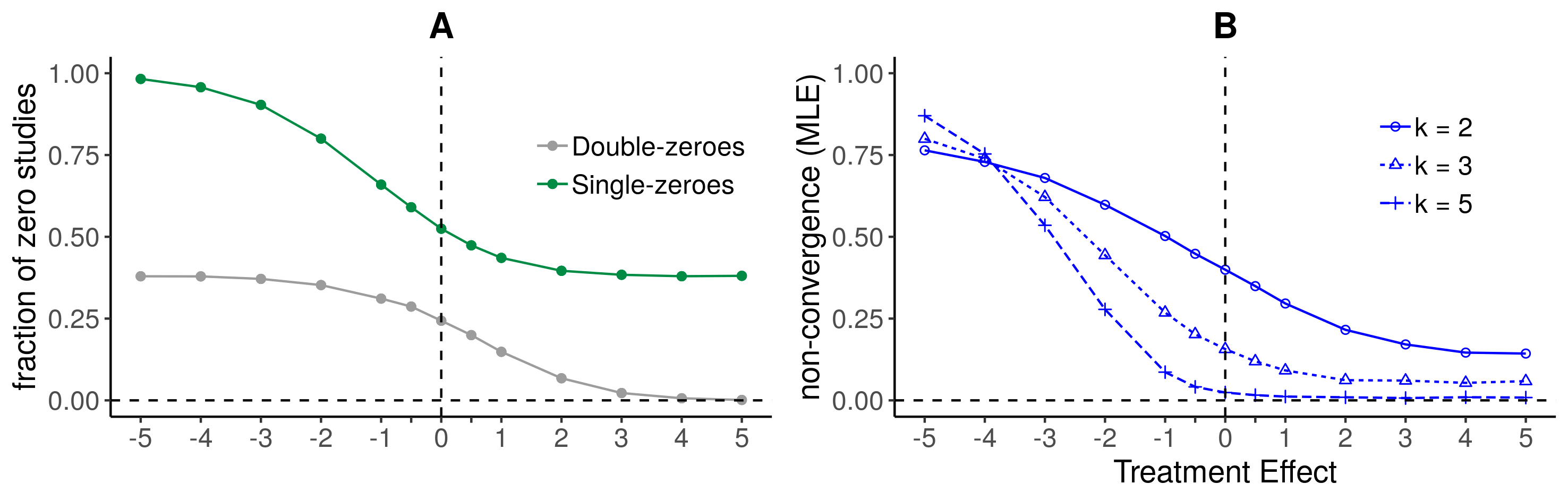}
 \caption{The average fraction of single-zero or double-zero studies in a simulated meta-analysis dataset (A), and the fraction of failed runs (non-convergence) for MLE with different numbers of studies $k$ used in the simulation (B) are shown.}
  \label{fig:SimConv}
\end{figure}

The proposed approach and two comparators are included in the analysis: BNHM using a vague prior ($\mathcal{N}(0, 100^2)$) for $\theta$ (``Vague''), BNHM using a WIP ($\mathcal{N}(0, 2.82^2)$) for $\theta$ (``WIP''), and BNHM using MLE (``MLE''). For both Vague and WIP approaches, the prior for $\tau$ and $\mu$ are taken as $\mathcal{HN}(0.5)$, and $\mathcal{N}(0, 10)$, respectively.  Estimation failure were assumed to be 0 for Bayesian methods, and convergence was assumed to be reached after \mbox{1\,000} iterations of burn-in. 3 chains were run in parallel for a total of \mbox{2\,000} iterations. with  We used the package \texttt{lme4} for the MLE, whereas the Vague and WIP methods were fitted with our \textbf{MetaStan} package. The highest density intervals were obtained using the \texttt{HDInterval}\cite{HDInterval} package. All computations were done in \textbf{R}\@.

For the MLE, the fraction of failed runs (non-convergence) is shown in Figure 3B. Estimation failure is closely related to the fraction of meta-analysis datasets including single-zero or double-zero studies in the dataset, which can be seen by comparing Figure 3A and Figure 3B. This is because when the data is highly sparse, estimation becomes more challenging for the MLE\@. As a performance measure, we use the bias ($\frac{1}{N} \sum_{i = 1}^{N}(\hat{\theta}-\theta)$) based on the MLE and posterior medians. Moreover, the coverage probability and the mean length of interval estimates for $\theta$ are reported.

%=================================================
\subsection{Simulation results} \label{sec:results}
%=================================================
\begin{figure}
  \centering
   \includegraphics[scale=0.55]{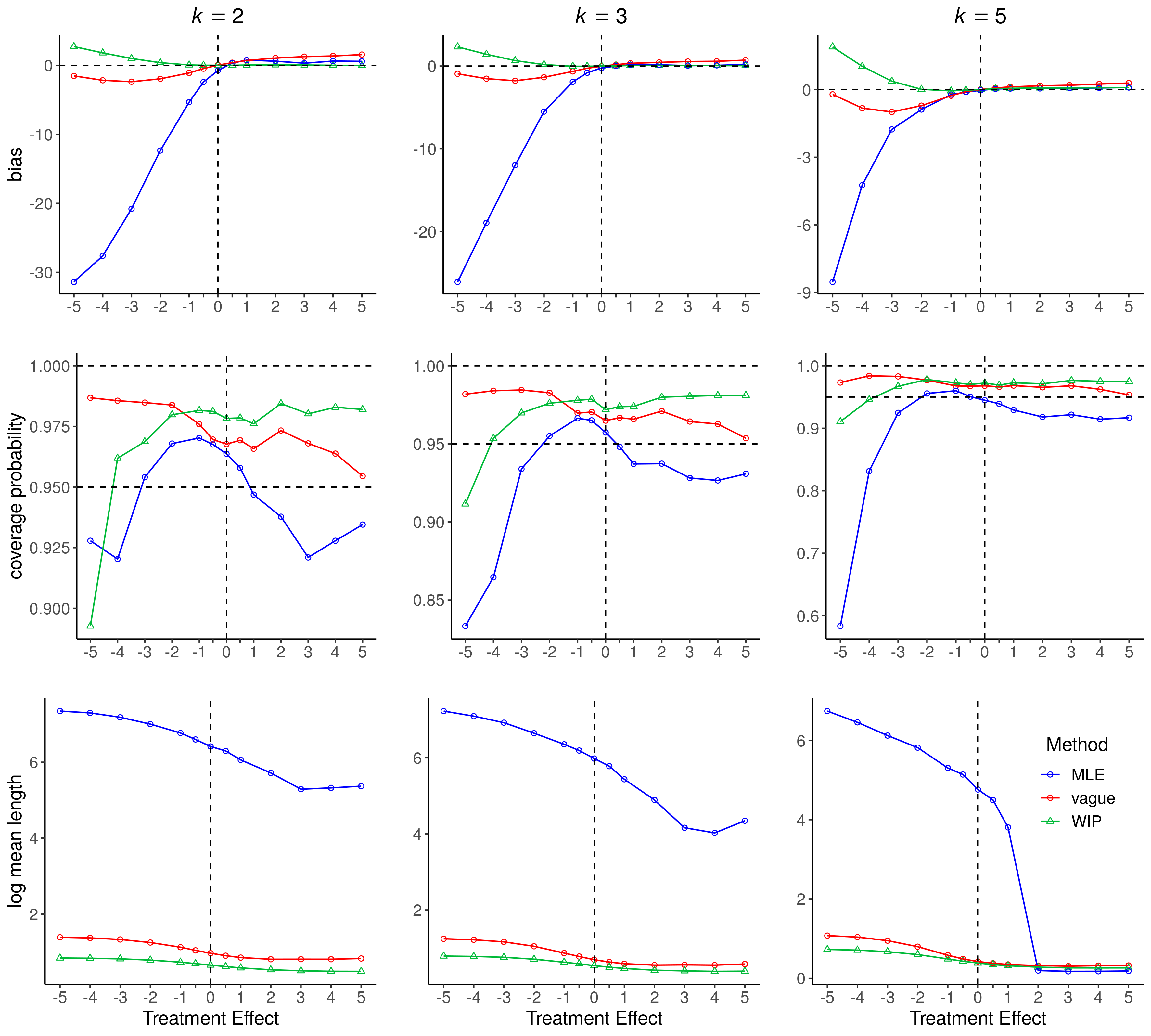}
 \caption{The bias for the mean treatment effect $\theta$, coverage probabilities and log mean length of the interval estimates for $\theta$ obtained by three methods (MLE, Vague and WIP) are shown.}
  \label{fig:SimRareBNHM1}
\end{figure}

The bias for posterior medians from the Vague and the WIP methods, and for maximum likelihood estimate from the  MLE across scenarios are displayed in the first row of Figure 4. Note that failed runs were excluded from the calculation of performance measures which is relevant only for the MLE\@. The MLE shows unacceptably high bias for the scenarios with $\theta \leq 0$, corresponding to the scenarios in which fraction of zero studies is very high. The WIP displays somewhat positive bias whereas the Vague shows negative bias for the scenarios with $\theta \leq 0$. It is important to note that the results of the bias behave similar to the fraction of zero studies and the fraction of non-convergence of the MLE, meaning that the bias is higher in scenarios with more sparse data. Since the Vague approach uses a vague prior on $\theta$, one would expect a somewhat similar behaviour of bias from the Vague and the MLE approaches. However, the fact the data is highly sparse and also the Vague approach includes a weakly informative prior for $\tau$ may be explanations of the better performance of the Vague method in comparison to the MLE\@. Moreover, performance in terms of bias is improving for all methods when the number of studies $k$ is increasing.

Figure 4 also shows coverage probabilities and log mean lengths for 95\% highest density intervals (HDIs) obtained by the Vague and the WIP, and for 95\% Wald confidence intervals (CIs) obtained by the MLE\@. The WIP method shows higher coverage than nominal level across all different true treatment effects except $\theta = -5$. On the other hand, the HDIs obtained by WIP are shorter in comparison to HDIs obtained by the Vague and CIs obtained by the MLE approaches.

\begin{figure}
  \centering
   \includegraphics[scale=0.55]{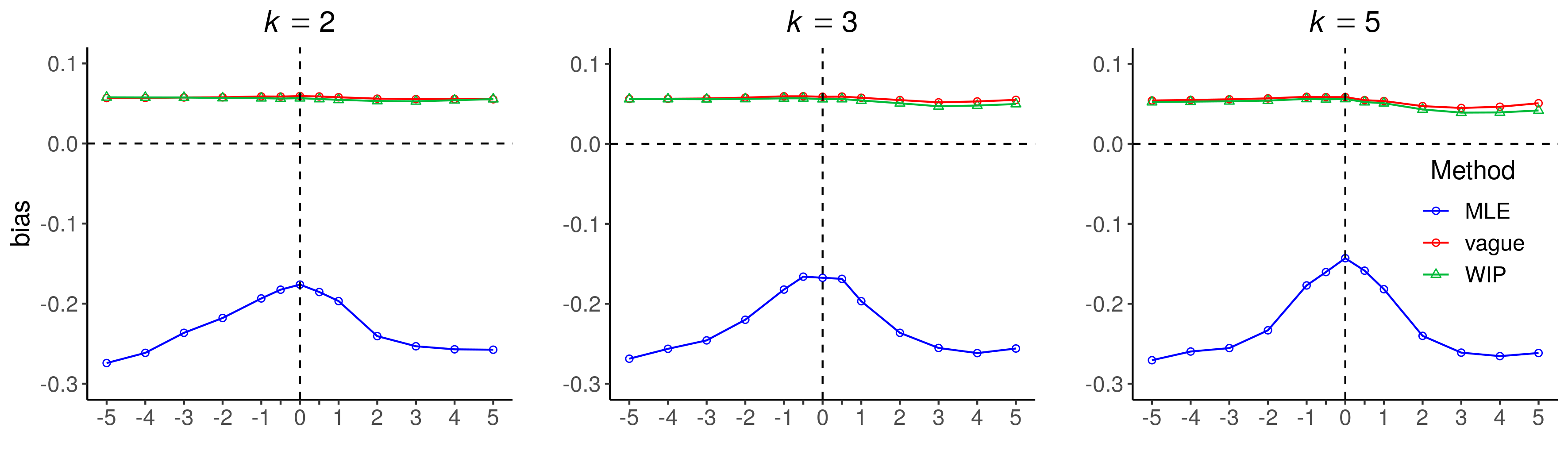}
 \caption{The bias for the heterogeneity parameter $\tau$ obtained by three methods (MLE, Vague and WIP) are shown. True heterogeneity standard deviation is assumed to be $\tau=0.28$.}
  \label{fig:SimRareBNHM1tau}
\end{figure}

Lastly, the bias for the heterogeneity parameter $\tau$ obtained by three methods (the MLE, the Vague and the WIP) are demonstrated in Figure 5. For Bayesian methods, posterior medians are used as the point estimates. Recall that, the prior used for $\tau$ in the Vague and the WIP is a weakly informative prior ($\mathcal{HN}(0.5)$). The MLE under-estimates the true heterogeneity, whereas the Vague and the WIP methods, slightly, over-estimates it. These observations are in aligned with the conclusions made by Friede et al.\cite{JRSM:JRSM1217}
 
%=================================================
\section{Example revisited} \label{sec:revisited}
%=================================================
Returning to the dataset described in Section~\ref{sec:app}, we consider the data on death and PTLD outcomes shown in Table~\ref{tab:Data}. The observed log odds ratios are displayed in Figure 6. To be able to visualize the observed log-odds ratios when there is single-zero or double-zero study, a continuity correction of 0.5 is added to all cells of the single-zero or double-zero study's contingency table. The wide CIs for observed log-odds ratios reflect the rather small sample sizes in the datasets. Furthermore, the variability in the point estimates can be seen as an indicator of the heterogeneity between trials.

We analyse the datasets using three methods investigated in the simulation studies, namely Vague, WIP, and MLE approaches. The code to implement the MLE is shown in Appendix~\ref{app2}. Recall that the only difference between Vague and WIP is the prior used for the treatment effect parameter $\theta$ in the model, namely $\mathcal{N}(0, 100^2)$ for the former and $\mathcal{N}(0, 2.82^2)$ for the latter. WIP can be implemented in a routine data analysis using our \texttt{MetaStan} package as follows
\begin{knitrout}
\definecolor{shadecolor}{rgb}{0.969, 0.969, 0.969}\color{fgcolor}\begin{kframe}
\begin{alltt}
\hlstd{bnhm.wip.CrinsPTLD.stan}  \hlkwb{<-} \hlkwd{meta_stan}\hlstd{(}\hlkwc{ntrt} \hlstd{= dat.Crins2014.PTLD}\hlopt{$}\hlstd{exp.total,}
                                      \hlkwc{nctrl} \hlstd{= dat.Crins2014.PTLD}\hlopt{$}\hlstd{cont.total,}
                                      \hlkwc{rtrt} \hlstd{= dat.Crins2014.PTLD}\hlopt{$}\hlstd{exp.PTLD.events,}
                                      \hlkwc{rctrl} \hlstd{= dat.Crins2014.PTLD}\hlopt{$}\hlstd{cont.PTLD.event,}
                                      \hlkwc{tau_prior_dist} \hlstd{=} \hlstr{"half-normal"}\hlstd{,}
                                      \hlkwc{tau_prior} \hlstd{=} \hlnum{0.5}\hlstd{,}
                                      \hlkwc{delta} \hlstd{=} \hlnum{250}\hlstd{)}
\end{alltt}
\end{kframe}
\end{knitrout}

The argument \texttt{delta} corresponds to $\delta$ from \eqref{eq:WIPprior}, and thus is used to calculate the WIP for $\theta$. Alternatively, one can directly specify the prior for $\theta$, in our case, equivalently, we can have \texttt{theta\_prior = c(0, 2.82)}. The Vague method is, simply, implemented by omitting the argument \texttt{delta\_u} and specifying \texttt{theta\_prior = c(0, 100)}. To check MCMC convergence, we use the Gelman-Rubin statistics, and traceplots. For the WIP approach, the corresponding traceplots are shown in Figure 7 and Figure 8 (Appendix~\ref{app1}) for death and PTLD outcomes, respectively. There was no divergence reported for both datasets. The MLE fit for the dataset with death outcome does not cause any warning from \texttt{lme4}. For PTLD outcomes, \texttt{lme4} gives a warning suggesting that the estimates may not be reliable. Nevertheless, it produces the MLE estimate and CI for treatment effect parameter, and we report them. The results from three methods are shown in Figure 6. For the MLE method, the MLE and 95\% CI are given. For Vague and WIP methods, posterior medians and 95\% HDIs are shown. The estimates of the between-trial heterogeneity $\hat{\tau}$ are also included in the figure. The point estimates from three methods look quite similar. MLE gives shorter interval estimates compared to Bayesian alternatives which is due to the fact that the uncertainty of $\tau$ is taken into account in both Bayesian approaches. In the original paper, Crins et al\cite{PETR:PETR12362} fitted a normal-normal hierarchical model using REML,\cite{metafor} and risk ratio was used as the measure of treatment effect. They concluded that treatment IL-2RA failed to show statistically significant result for reducing death. We obtained similar point estimates with somewhat wider interval estimates to Crins et al,\cite{PETR:PETR12362} specifically their risk ratio estimate was 0.61 (CI 0.27-1.37), and we obtained the odds ratio estimate 0.57 (HDI 0.21-1.46) using the WIP method. Concerning the PTLD, the risk ratio was estimated as 1.60 (CI 0.20-12.67) by Crins et al,\cite{PETR:PETR12362} the odds ratio is estimated 1.99 (HDI 0.20-25.35) using the WIP method. The wider interval estimates obtained by WIP may stem from the fact that the uncertainty of $\tau$ is taken into account. 

Considering death as outcomes, the heterogeneity parameter $\tau$ is estimated 0.30, 0.28, and 0.00 using WIP, Vague, and MLE, respectively. Similarly for the PTLD outcomes, for $\hat{\tau}$, we obtained 0.33, 0.33, and 0.00 using WIP, Vague, and MLE, respectively. Moreover, Crins et al\cite{PETR:PETR12362} concluded that there is no evidence for heterogeneity between trials using using Cochrane's Q test for both death and PTLD outcomes. Since the prior used for $\tau$ is the same for WIP and Vague, similar heterogeneity estimates are expected. On the other hand, the MLE estimate ($\hat{\tau} = 0.00$) is most probably underestimating the actual amount of heterogeneity.

\begin{figure}
  \centering
  \includegraphics[scale=0.65]{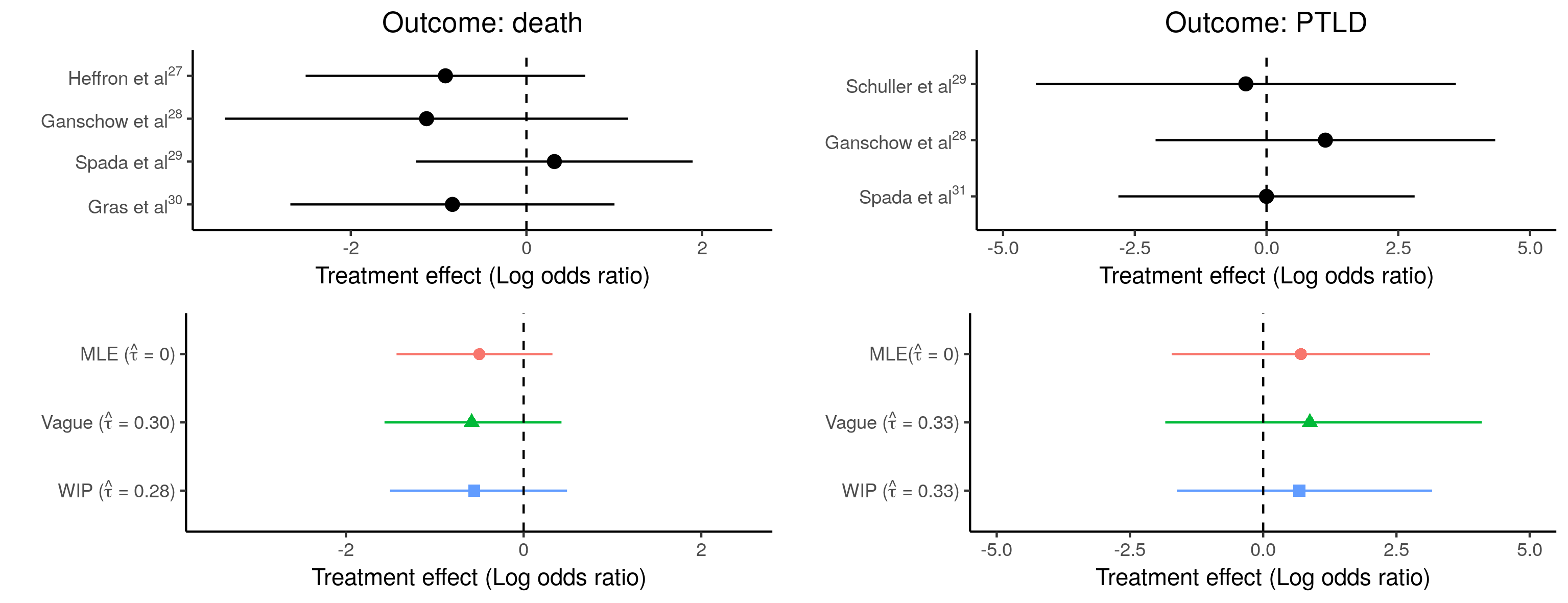}
  \caption{The motivating paediatric transplantation application: Top figures are the observed log-odds ratios for death and PTLD outcomes (computed using a continuity correction in case of zero counts). The bottom two panels show mean treatment effect estimates of $\theta$ obtained by Vague, WIP, and MLE\@. Heterogeneity parameter estimates $\hat{\tau}$ are also given on the left.}
  \label{fig:CrinsRes}
\end{figure}

%=================================================
\section{Conclusions and Discussion} \label{sec:discuss}
%=================================================
Random effects meta-analyses are commonly based on few studies only, which often poses problems for inference, as certain asymptotics cannot be relied upon. Additional issues arise for binary outcomes when only few or no events are observed in some of the studies or study arms. To deal with such data sparsity in the meta-analysis, we have proposed the use of weakly informative priors (WIPs) for the treatment effect parameter $\theta$ in a binomial-normal hierarchical model (BNHM). We demonstrated how a normal WIP for $\theta$ can be derived by considering an a priori interval for the treatment effect on a log odds ratio scale. Also, the empirical evidence obtained from \mbox{$37\,773$} meta-analyses with binomial outcomes from the Cochrane Database of Systematic Reviews supports the proposed WIP\@. In simulation studies, the suggested method displays lower bias for $\theta$ and substantially shorter interval estimates for $\theta$ with somewhat higher coverage than nominal level in comparison to alternative methods.

The simulation results displayed in Figure 4 are somewhat in contrast to the results given by Friede et al,\cite{JRSM:JRSM1217} who observed lower coverage than nominal level of MLE-methods in a similar setting, but not based on rare events. We also investigated a scenario closer to their setup by considering higher baseline risks between 0.05 and 0.20. The results are shown in Figure 9 in the Appendix~\ref{app3}, and indeed here the MLE method exhibits lower coverage than nominal level, as reported by Chung et al\cite{SIM:SIM5821} and Friede et al.\cite{JRSM:JRSM1217} The high bias and too wide interval estimates obtained by the Vague and the MLE are still present, but not as high as in the results of the simulations in which true baseline risks are relatively lower.

The introduced approach is not restricted to the BNHM as introduced above; similar approaches may analogously be defined in similar cases, e.g. a Poisson-normal hierarchical model. However, a crucial point is that the treatment effect parameter is explicitly parameterized in the model, so that it can directly be ``penalized'' via the prior specification. Hence, so-called contrast-based models\cite{dias2015absolute} (in which relative treatment effects are assumed to be exchangeable across trials) are suitable for this purpose unlike arm-based models.

Jackson et al\cite{SIM:SIM7588} investigated 7 random effects meta-analysis models including the BNHM which we consider in this paper (``Model 4'' in Jackson et al\cite{SIM:SIM7588}) and another parametrization of the BNHM (``Model 2'' in Jackson et al\cite{SIM:SIM7588}). The only difference between two models is that in their Model 2, the treatment indicator $x_{ik}$ of \eqref{eq:BNHM} is +1 for the experimental arm, and 0 for the control arm. Note that commonly used network meta-analysis models for example\cite{dias2011glmm} are generalizations of Model 2 in Jackson et al.\cite{SIM:SIM7588} As reported by Jackson et al,\cite{SIM:SIM7588} we also observe the underestimation of the heterogeneity parameter $\tau$, and hence decided to only consider their Model 4. On the other hand, it is important to note that the usage of a weakly informative prior for $\theta$ also improves the performance in Model 2, as we have seen for the Model 4.

The BNHM can be extended to a network meta-analysis model\cite{doi:10.1002/jrsm.1285} which is desirable if there are multiple treatments, and/or multi-arm trials in the dataset. Even if the data set in a network meta-analysis consists of many studies overall, some of the treatment effects may still be informed by few studies only. Thus, the use of WIPs for treatment effect parameters in the context of network meta-analysis with rare events can be very helpful. One may find it restrictive to have a normal prior for $\theta$, it may be worth exploring alternatives like Cauchy or log-$F$ distributions\cite{gelman2008, greenland2015penalization} for penalization. Different distributions as WIP for $\theta$, different parametrizations of BNHM, or different data models can be implemented in \textbf{Stan} or MCMC methods in general. Although, currently, our package \texttt{MetaStan} is restricted to use a BNHM for pairwise meta-analysis, it can be extended to conduct meta-analysis and network meta-analysis with flexible data model and prior options. 

%=================================================
\section*{Acknowledgements}
%=================================================
We thank Leonhard Held who contributed valuable comments and pointed us to several important references.

\appendix

%==================================================================
\section{How to use \texttt{MetaStan} \textbf{R} package? \label{app1}}
%===================================================================
The development version of \textbf{MetaStan} is available on Github (\href{https://github.com/gunhanb/MetaStan}{https://github.com/gunhanb/MetaStan}) and can be installed using \texttt{devtools} package as follows:

\begin{knitrout}
\definecolor{shadecolor}{rgb}{0.969, 0.969, 0.969}\color{fgcolor}\begin{kframe}
\begin{alltt}
\hlkwd{library}\hlstd{(devtools)}
\hlkwd{install_github}\hlstd{(}\hlstr{"gunhanb/MetaStan"}\hlstd{)}
\end{alltt}
\end{kframe}
\end{knitrout}

The example described in the text (Crins dataset) is available in the package, and it can be loaded as follows:

\begin{knitrout}
\definecolor{shadecolor}{rgb}{0.969, 0.969, 0.969}\color{fgcolor}\begin{kframe}
\begin{alltt}
\hlkwd{library}\hlstd{(}\hlstr{"MetaStan"}\hlstd{)}
\hlkwd{data}\hlstd{(}\hlstr{"dat.Crins2014"}\hlstd{,} \hlkwc{package} \hlstd{=} \hlstr{"MetaStan"}\hlstd{)}
\hlcom{## Subset of dataset where PTLD outcomes available}
\hlstd{dat.Crins2014.PTLD} \hlkwb{=} \hlkwd{subset}\hlstd{(dat.Crins2014,} \hlkwd{is.finite}\hlstd{(exp.PTLD.events))}
\end{alltt}
\end{kframe}
\end{knitrout}

Additional information can be obtained by typing \texttt{?dat.Crins2014} (for any dataset and function in the package). 

\texttt{meta\_stan} is the main fitting function of this package. The main computations are executed via the \textbf{rstan} package's \texttt{sampling} function. We can fit the binomial-normal hierarchical using a weakly informative prior for treatment effect as follows:
\begin{knitrout}
\definecolor{shadecolor}{rgb}{0.969, 0.969, 0.969}\color{fgcolor}\begin{kframe}
\begin{alltt}
\hlstd{bnhm.wip.CrinsPTLD.stan}  \hlkwb{<-} \hlkwd{meta_stan}\hlstd{(}\hlkwc{ntrt} \hlstd{= dat.Crins2014.PTLD}\hlopt{$}\hlstd{exp.total,}
                                      \hlkwc{nctrl} \hlstd{= dat.Crins2014.PTLD}\hlopt{$}\hlstd{cont.total,}
                                      \hlkwc{rtrt} \hlstd{= dat.Crins2014.PTLD}\hlopt{$}\hlstd{exp.PTLD.events,}
                                      \hlkwc{rctrl} \hlstd{= dat.Crins2014.PTLD}\hlopt{$}\hlstd{cont.PTLD.event,}
                                      \hlkwc{tau_prior_dist} \hlstd{=} \hlstr{"half-normal"}\hlstd{,}
                                      \hlkwc{tau_prior} \hlstd{=} \hlnum{0.5}\hlstd{,}
                                      \hlkwc{delta} \hlstd{=} \hlnum{250}\hlstd{,}
                                      \hlkwc{chains} \hlstd{=} \hlnum{4}\hlstd{,}
                                      \hlkwc{iter} \hlstd{=} \hlnum{2000}\hlstd{,}
                                      \hlkwc{warmup} \hlstd{=} \hlnum{1000}\hlstd{)}
\end{alltt}
\end{kframe}
\end{knitrout}

Convergence diagnostics and the results can be, very conveniently, obtained using \texttt{shinystan} package as follows:

\begin{knitrout}
\definecolor{shadecolor}{rgb}{0.969, 0.969, 0.969}\color{fgcolor}\begin{kframe}
\begin{alltt}
\hlkwd{library}\hlstd{(}\hlstr{"shinystan"}\hlstd{)}
\hlcom{## Firstly convert "stan" object to a "shinystan" object}
\hlstd{bnhm.wip.CrinsPTLD.shinystan} \hlkwb{=} \hlkwd{as.shinystan}\hlstd{(bnhm.wip.CrinsPTLD.stan}\hlopt{$}\hlstd{fit)}
\hlkwd{launch_shinystan}\hlstd{(bnhm.wip.CrinsPTLD.shinystan)}
\end{alltt}
\end{kframe}
\end{knitrout}

Traceplots for the estimated parameters $\theta$ and $\tau$ including burn-in are shown in Figure 7 and Figure 8 for death and PTLD outcomes, respectively.

Lastly, the posterior summary statistics can be obtained using the following command:
\begin{knitrout}
\definecolor{shadecolor}{rgb}{0.969, 0.969, 0.969}\color{fgcolor}\begin{kframe}
\begin{alltt}
\hlstd{bnhm.wip.CrinsPTLD.stan}\hlopt{$}\hlstd{fit_sum}
\end{alltt}
\end{kframe}
\end{knitrout}

\begin{figure}
  \centering
  \includegraphics[scale=0.55]{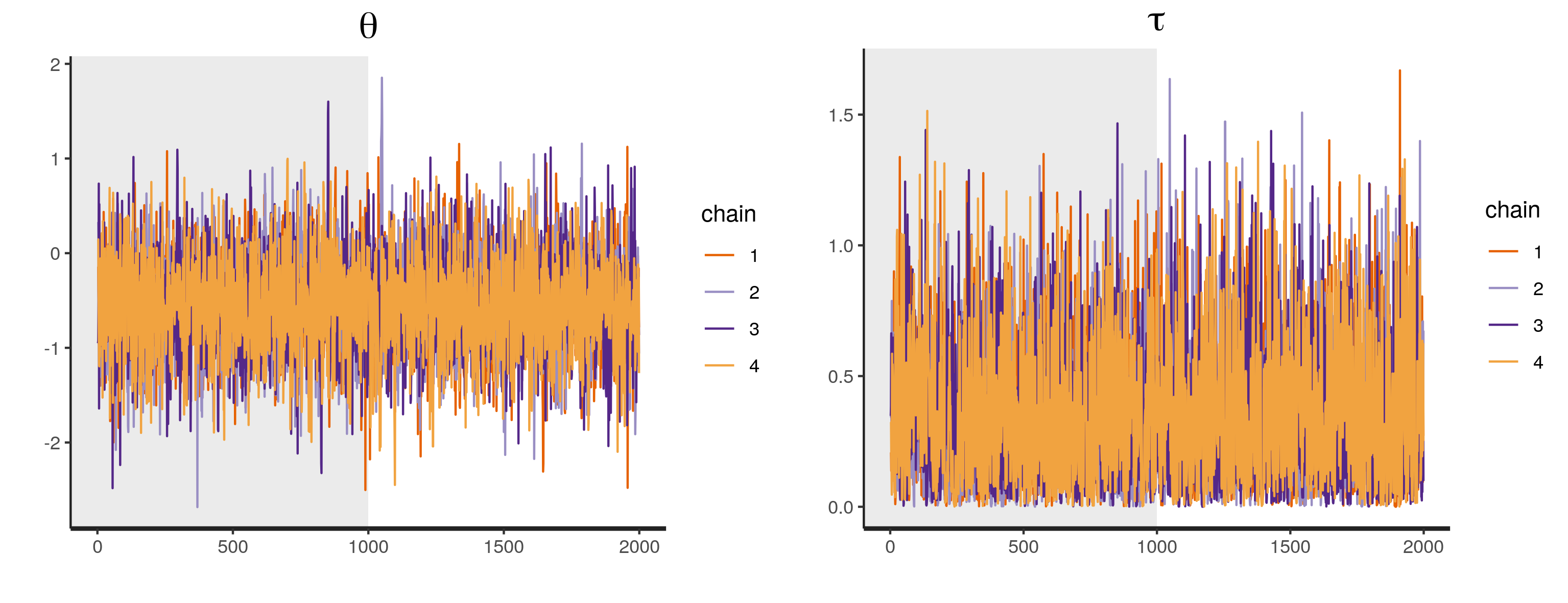}
  \caption{Traceplots for the estimated parameters $\theta$ and $\tau$ including burn-in for death outcomes.}
  \label{fig:Trace_death}
\end{figure}

\begin{figure}
  \centering
  \includegraphics[scale=0.55]{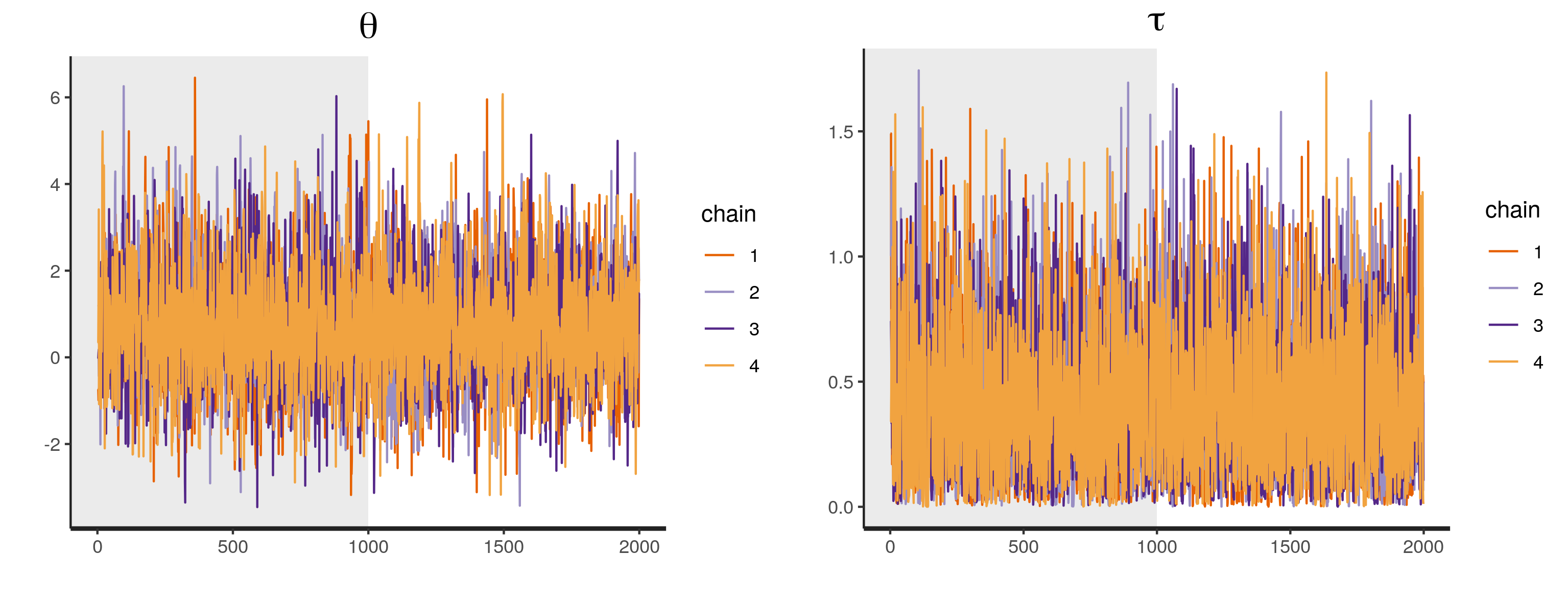}
  \caption{Traceplots for the estimated parameters $\theta$ and $\tau$ including burn-in for PTLD outcomes.}
  \label{fig:Trace_PTLD}
\end{figure}

%==================================================================
\section{\textbf{R} code to implement BNHM using the MLE method \label{app2}}
%===================================================================

\begin{knitrout}
\definecolor{shadecolor}{rgb}{0.969, 0.969, 0.969}\color{fgcolor}\begin{kframe}
\begin{alltt}
\hlkwd{library}\hlstd{(}\hlstr{"lme4"}\hlstd{)}
\hlcom{## Firstly convert dataset to a long format}
\hlcom{## using MetaStan:::convert_data_arm function}
\hlstd{dat.Crins2014.PTLD.long} \hlkwb{<-} \hlkwd{convert_data_arm}\hlstd{(dat.Crins2014.PTLD}\hlopt{$}\hlstd{exp.total,}
                                            \hlstd{dat.Crins2014.PTLD}\hlopt{$}\hlstd{cont.total,}
                                            \hlstd{dat.Crins2014.PTLD}\hlopt{$}\hlstd{exp.PTLD.events,}
                                            \hlstd{dat.Crins2014.PTLD}\hlopt{$}\hlstd{cont.PTLD.events)}

\hlkwd{glmer}\hlstd{(}\hlkwd{cbind}\hlstd{(r, sampleSize} \hlopt{-} \hlstd{r)} \hlopt{~} \hlkwd{factor}\hlstd{(mu)} \hlopt{+} \hlkwd{factor}\hlstd{(theta)} \hlopt{+} \hlstd{(theta12} \hlopt{-} \hlnum{1}\hlopt{|}\hlstd{mu),}
      \hlkwc{data} \hlstd{= dat.Crins2014.PTLD.long,} \hlkwc{family} \hlstd{=} \hlkwd{binomial}\hlstd{(}\hlkwc{link} \hlstd{=} \hlstr{"logit"}\hlstd{),} \hlkwc{nAGQ} \hlstd{=} \hlnum{7}\hlstd{)}
\end{alltt}
\end{kframe}
\end{knitrout}

% =======================================
\section{Additional simulation results \label{app3}} 
% =======================================
We also conducted simulations using the same settings as described in Section~\ref{sec:sim} under BNHM, but using higher baseline risk probabilities, specifically, baseline risks ($\mu_{i}$) are uniformly taken between 0.05 and 0.2. Results are illustrated in Figure 9 (analogous to Figure 5).

\begin{figure}
  \centering
   \includegraphics[scale=0.55]{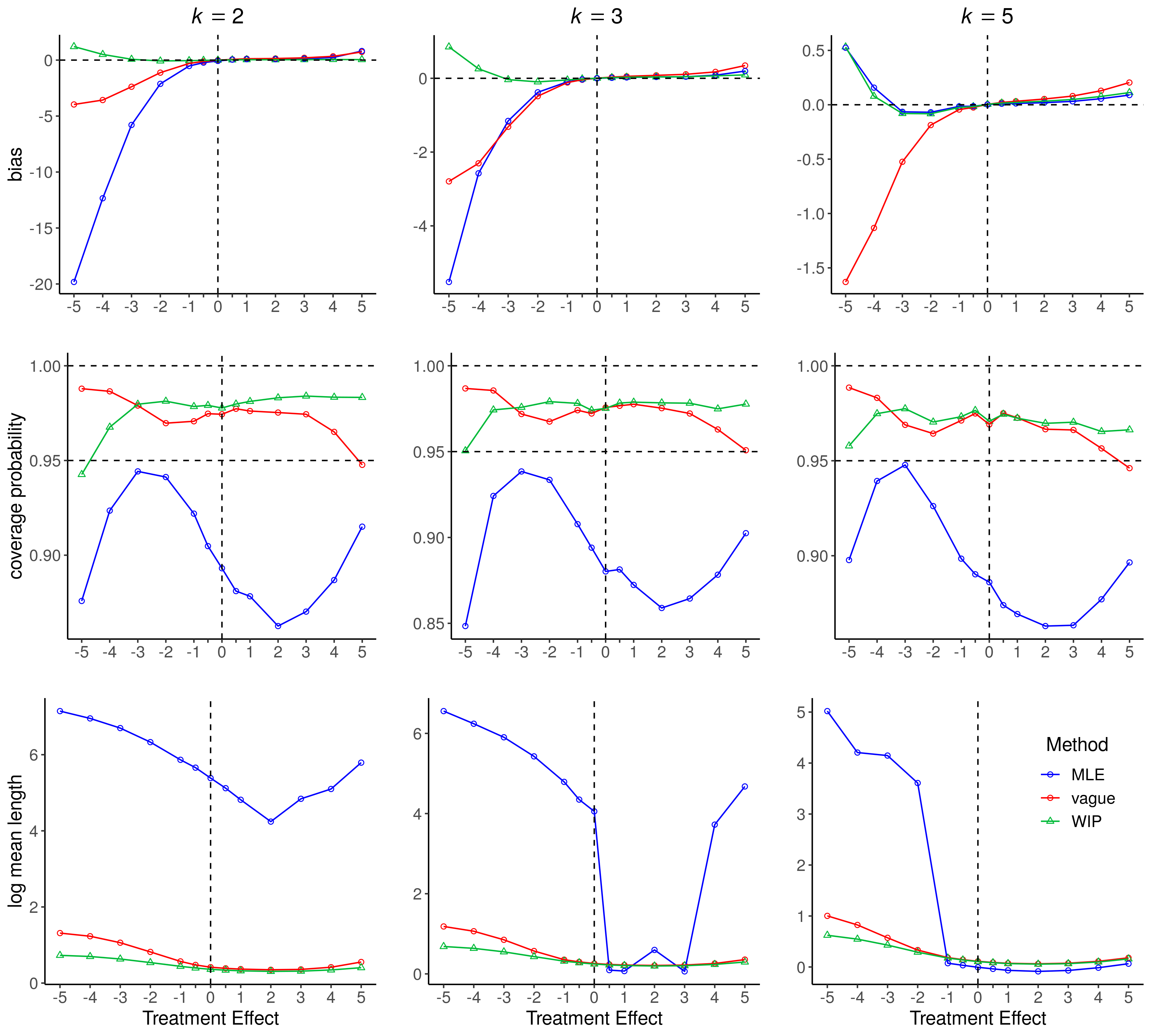}
 \caption{Simulations with high baseline risks: The bias for the mean treatment effect $\theta$, coverage probabilities and log mean length of the interval estimates for $\theta$ obtained by three methods (MLE, Vague and WIP) are shown.}
  \label{fig:SimNONrareBNHM1}
\end{figure}

% =======================================
\section*{Highlights} 
% =======================================
\noindent \textbf{What is already known}: Standard meta-analysis methods are not suitable for meta-analysis of few studies with rare events.

\noindent \textbf{What is new}: To deal with data sparsity present in the meta-analysis of few studies with rare events, we suggest the use of weakly informative priors as penalization for the treatment effect parameter.

\noindent \textbf{Potential impact for RSM readers outside the authors' field}: To make it more accessible
to meta-analyzers, a publicly available \textbf{R} package, \texttt{MetaStan}, is developed for fitting Bayesian meta-analysis models using weakly informative priors.

\bibliography{references}

\begin{thebibliography}{10}

\bibitem{higginscochrane}
Higgins JPT, Green S~(eds). {\it Cochrane Handbook for Systematic Reviews of
  Interventions.}
\newblock Chichester: Wiley; 2008.

\bibitem{Hedges1985189}
Hedges LV, Olkin I.~Random Effects Models for Effect~Sizes. {\it Statistical
  Methods for Meta-Analysis}.
\newblock San Diego: Academic Press; 1985.

\bibitem{SIM:SIM2528}
Bradburn MJ, Deeks JJ, Berlin JA, Russell~Localio A. Much ado about nothing: a
  comparison of the performance of meta-analytical methods with rare events.
  {\it Stat Med. } 2007;26:53--77.

\bibitem{SIM:SIM7588}
Jackson D, Law M, Stijnen T, Viechtbauer W, White IR. A comparison of seven
  random-effects models for meta-analyses that estimate the summary odds ratio.
   {\it Stat Med. } 2018;37:1059--1085.

\bibitem{SIM:SIM4040}
Stijnen T, Hamza TH, \"Ozdemir P. Random effects meta-analysis of event outcome
  in the framework of the generalized linear mixed model with applications in
  sparse data.  {\it Stat Med. } 2010;29:3046--3067.

\bibitem{BIMJ:BIMJ1584}
B\"ohning D, Mylona K, Kimber A. Meta-analysis of clinical trials with rare
  events.  {\it Biom J. } 2015;57:633--648.

\bibitem{SIM:SIM3964}
Cai T, Parast L, Ryan L. Meta-analysis for rare events.  {\it Stat Med. }
  2010;29:2078--2089.

\bibitem{SIM:SIM6383}
Kuss O. Statistical methods for meta-analyses including information from
  studies without any events--add nothing to nothing and succeed nevertheless.
  {\it Stat Med. } 2015;34:1097--1116.

\bibitem{SIM:SIM4780142408}
Smith TC, Spiegelhalter DJ, Thomas A. Bayesian approaches to random-effects
  meta-analysis: A comparative study.  {\it Stat Med. } 1995;14:2685--2699.

\bibitem{albert1984}
Albert A, Anderson JA. On the existence of maximum likelihood estimates in
  logistic regression models.  {\it Biometrika. } 1984;71:1--10.

\bibitem{greenland2015penalization}
Greenland S, Mansournia MA. Penalization, bias reduction, and default priors in
  logistic and related categorical and survival regressions.  {\it Stat Med. }
  2015;34:3133--3143.

\bibitem{Firth1993}
Firth D. Bias reduction of maximum likelihood estimates.  {\it Biometrika. }
  1993;80:27--38.

\bibitem{Heinze2002}
Heinze G, Schemper M. A solution to the problem of separation in logistic
  regression.  {\it Stat Med. };21:2409--2419.

\bibitem{gelman2008}
Gelman A, Jakulin A, Pittau MG, Su~Y. A weakly informative default prior
  distribution for logistic and other regression models.  {\it Ann Appl Stat. }
  2008;2:1360-1383.

\bibitem{Davey2011}
Davey J, Turner RM, Clarke MJ, Higgins JPT. Characteristics of meta-analyses
  and their component studies in the Cochrane Database of Systematic Reviews: a
  cross-sectional, descriptive analysis.  {\it BMC Med Res Methodol. }
  2011;11:160.

\bibitem{SIM:SIM5821}
Chung Y, Rabe-Hesketh S, Choi IH. Avoiding zero between-study variance
  estimates in random-effects meta-analysis.  {\it Stat Med. }
  2013;32:4071--4089.

\bibitem{gelman2006}
Gelman A. Prior distributions for variance parameters in hierarchical models
  (comment on article by Browne and Draper).  {\it Bayesian Anal. }
  2006;1:515--534.

\bibitem{JRSM:JRSM1217}
Friede T, R\"over C, Wandel S, Neuenschwander B. Meta-analysis of few small
  studies in orphan diseases.  {\it Res Synth Methods. } 2017;8:79--91.

\bibitem{BIMJ:BIMJ1725}
Friede T, R\"over C, Wandel S, Neuenschwander B. Meta-analysis of two studies
  in the presence of heterogeneity with applications in rare diseases.  {\it
  Biom J. } 2017;59:658--671.

\bibitem{williams2018bayesian}
Williams, DR and Rast, P and B\"urkner, P. Bayesian meta-analysis with weakly
  informative prior distributions. PsyarXiv Preprints.
  https://psyarxiv.com/7tbrm. Updated July 02, 2018. Accessed August 24, 2018.

\bibitem{Benderetal2018}
Bender R, Friede T, Koch A, et al. Methods for evidence synthesis in the case
  of very few studies.  {\it Res Synth Methods. } 2018:1--11.

\bibitem{dias2015absolute}
Dias S, Ades AE. Absolute or relative effects? {A}rm-based synthesis of trial
  data.  {\it Res Synth Methods. } 2016;7:23-28.

\bibitem{PETR:PETR12362}
Crins ND, R\"over C, Goralczyk AD, Friede T. Interleukin-2 reeptor antagonists
  for pediatric liver transplant recipients: A systematic review and
  meta-analysis of controlled studies.  {\it Pediatr Transplant. }
  2014;18:839--850.

\bibitem{metafor}
Viechtbauer W. Conducting meta-analyses in {R} with the {metafor} package.
  {\it J Stat Softw. } 2010;36:1--48.

\bibitem{Higgins557}
Higgins JPT, Thompson SG, Deeks JJ, Altman DG. Measuring inconsistency in
  meta-analyses.  {\it BMJ. } 2003;327:557--560.

\bibitem{SCHULLER20051151}
Schuller S, Wiederkehr JC, Coelho-Lemos IM, Avilla SG, Schultz C. Daclizumab
  induction therapy associated with Tacrolimus-MMF has better outcome compared
  with Tacrolimus-MMF alone in pediatric living donor liver transplantation.
  {\it Transplant Proc. } 2005;37:1151--1152.

\bibitem{AJT:AJT1406}
Spada M, Petz W, Bertani A, et al. Randomized trial of basiliximab induction
  versus steroid therapy in pediatric liver allograft recipients under
  tacrolimus Iimmunosuppression.  {\it Am J Transplant. } 2006;6:1913--1921.

\bibitem{PETR:PETR371}
Ganschow R, Grabhorn E, Schulz A, Hugo AV, Rogiers X, Burdelski M. Long-term
  results of basiliximab induction immunosuppression in pediatric liver
  transplant recipients.  {\it Pediatr Transplant. } 2005;9:741--745.

\bibitem{Spiegelhalter2004bayesian}
Spiegelhalter DJ, Abrams KR, Myles JP. Prior~Distributions. {\it Bayesian
  Approaches to Clinical Trials and Health-Care Evaluation.}
\newblock West Sussex: CRC Press; 2004.

\bibitem{Greenland01062006}
Greenland S. Bayesian perspectives for epidemiological research: I. Foundations
  and basic methods.  {\it Int J Epidemiol. } 2006;35:765--775.

\bibitem{doi:10.1080/01621459.1995.10476592}
Kass RE, Wasserman L. A reference Bayesian test for nested hypotheses and its
  relationship to the Schwarz criterion.  {\it J Am Stat Assoc. }
  1995;90:928--934.

\bibitem{bayesmeta}
R\"over, C. Bayesian random-effects meta-analysis using the bayesmeta R
  package. arXiv.org E-print Archive. https://arxiv.org/abs/1711.08683. Updated
  November 23, 2017. Accessed August 30, 2018.

\bibitem{cochrane}
The Cochrane. Cochrane Database of Systematic Reviews.
  https://www.cochranelibrary.com. Accessed April 15, 2018.

\bibitem{davidaspringate201410782}
Springate, D. Cochrane\_scraper v1.1.0. https://doi.org/10.5281/zenodo.10782.
  Updated July, 2014. Accessed April, 2018.

\bibitem{doi:10.1093/ije/dys041}
Turner RM, Davey J, Clarke MJ, Thompson SG, Higgins JPT. Predicting the extent
  of heterogeneity in meta-analysis, using empirical data from the Cochrane
  Database of Systematic Reviews.  {\it Int J Epidemiol. } 2012;41:818-827.

\bibitem{10.1371/journal.pone.0069930}
Kontopantelis E, Springate DA, Reeves D. A re-Analysis of the Cochrane Library
  data: The dangers of unobserved heterogeneity in meta-analyses.  {\it PLoS
  One. } 2013;8:1--14.

\bibitem{lme4}
Bates D, M{\"a}chler M, Bolker B, Walker S. Fitting linear mixed-effects models
  using lme4.  {\it J Stat Softw. } 2015;67:1--48.

\bibitem{stan}
{Stan Development Team}. Stan Modeling Language User's Guide and Reference
  Manual. Version 2.17.0. 2017.

\bibitem{HMC}
Betancourt, M. A conceptual introduction to Hamiltonian Monte Carlo. arXiv.org
  E-print Archive. https://arxiv.org/abs/1701.02434. Updated July 16, 2018.
  Accessed August 24, 2018.

\bibitem{betancourt2015hamiltonian}
Betancourt M, Girolami M.~Hamiltonian {M}onte {C}arlo for hierarchical~models.
  {\it Current trends in {B}ayesian methodology with applications.}
\newblock Boca Raton: CRC Press; 2015.

\bibitem{HDInterval}
Meredith, M and Kruschke, J. HDInterval: Highest (Posterior) Density Intervals.
  R package version 0.2.0. 2018. https://CRAN.R-project.org/package=HDInterval

\bibitem{dias2011glmm}
Dias S, Welton NJ, Sutton AJ, Ades AE. {NICE} {DSU} {T}echnical support
  document 2: {A} generalised linear modelling framework for pairwise and
  network meta-analysis of randomised controlled trials.   2011.
\newblock last updated September 2016.

\bibitem{doi:10.1002/jrsm.1285}
G\"unhan BK, Friede T, Held L. A design-by-treatment interaction model for
  network meta-analysis and meta-regression with integrated nested Laplace
  approximations.  {\it Res Synth Methods. } 2018;9:179--194.

\bibitem{heffron2003}
Heffron TG, Pillen T, Smallwood GA, Welch D, Oakley B, Romero R. Pediatric
  liver transplantation with daclizumab induction therapy.  {\it
  Transplantation. } 2003;75:2040--2043.

\bibitem{LT:LT21397}
Gras JM, Gerkens S, Beguin C, et al. Steroid-free, tacrolimus-basiliximab
  immunosuppression in pediatric liver transplantation: Clinical and
  pharmacoeconomic study in 50 children.  {\it Liver Transpl. }
  2008;14:469--477.

\end{thebibliography}

\clearpage

\end{document}